\documentclass[aps,prl,groupedaddress,notitlepage]{revtex4-2}
\usepackage{fullpage}
\usepackage{graphicx} 
\usepackage{amsmath}
\usepackage{amsfonts}
\usepackage{amssymb}
\usepackage{hyperref}
\usepackage{float}
\usepackage{mdframed}
\usepackage{color}
\usepackage[T1]{fontenc} 

\usepackage{cleveref}

\usepackage{tikz}
\usetikzlibrary{shapes,arrows,chains}

\definecolor{explanationcolor}{RGB}{240,240,255}
\definecolor{examplecolor}{RGB}{250,240,240}
\definecolor{problemcolor}{RGB}{225,255,240}
\definecolor{challengecolor}{RGB}{255,240,225}
\definecolor{relevlitcolor}{RGB}{250,250,250}
\definecolor{hyperlinkcolor}{RGB}{0,0,0}
\definecolor{hypercitecolor}{RGB}{0,180,90}
\usepackage{hyperref}

\colorlet{lcfree}{black}
\colorlet{lcnorm}{black}
\colorlet{lccong}{black}

\newenvironment{explanationbox}{\begin{mdframed}[backgroundcolor=explanationcolor,linewidth=1pt]}{\end{mdframed}}
\newenvironment{examplebox}{\begin{mdframed}[backgroundcolor=examplecolor,linewidth=0pt]}{\end{mdframed}}

\newcounter{explanation}
\def\theexplanation{\arabic{explanation}}

\newcounter{example}
\def\theexample{\arabic{example}}
\newenvironment{example}[1][]{\begin{examplebox}\refstepcounter{example}\par\medskip
   \noindent \textbf{Example~\theexample:} #1}{\medskip\end{examplebox}}

\newcounter{problem}
\def\theproblem{\arabic{problem}}
\newenvironment{problem}[1][]{\refstepcounter{problem}\par\medskip
   \noindent \textbf{Problem~\theproblem:} #1}{\medskip}
    
\begin{document}

\author{Kevin H\"ollring}
\affiliation{Interdisciplinary center for nanostructured films \& Institute for Theoretical Physics, Friedrich-Alexander-Universität Erlangen-Nürnberg, Germany}
\author{Ana-Sun\v{c}ana Smith}
\affiliation{Interdisciplinary center for nanostructured films \& Institute for Theoretical Physics, Friedrich-Alexander-Universität Erlangen-Nürnberg, Germany}
\affiliation{Division of Physical chemistry, Ru{\dj}er Boskovic Institute, Croatia}


\title{Mesoscopic modelling of epithelial tissues}

\maketitle


\begin{figure}[htp]
    \centering
    \includegraphics[width=.9\textwidth]{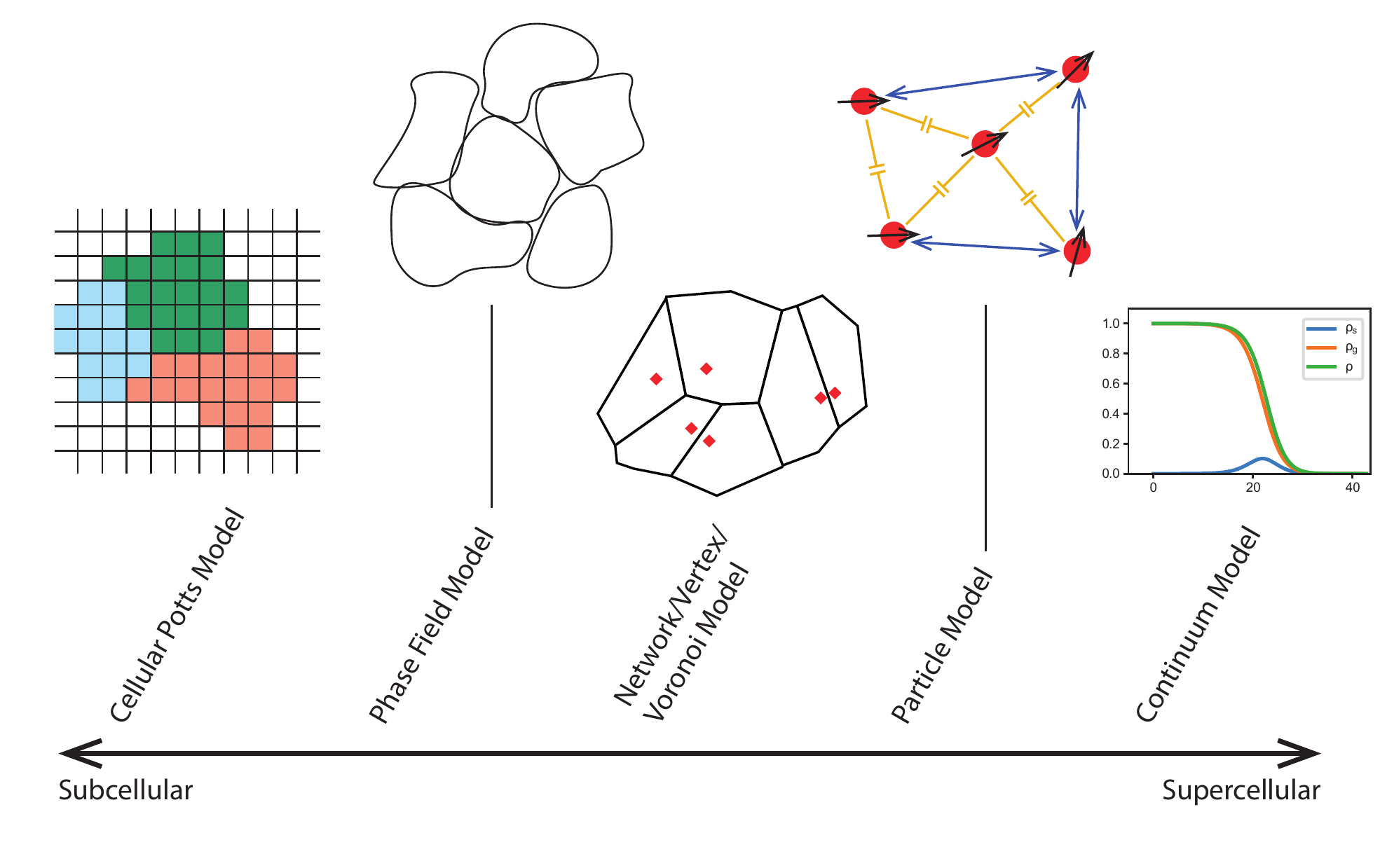}
    \caption{
    \textbf{Overview of cell models at different levels of detail and granularity.} 
        Starting from the Cellular Potts and Phase Field models at high level of detail, the Active Network models take cell descriptions towards a more abstract representation of shapes with fewer degrees of freedom.
        The Particle Models do not consider cell shape at all, only accounting for pairwise interaction forces of point-representations with the Continuum models finally abstracting away the individual cell altogether.}
    \label{fig:VM:model_overview}
\end{figure}

\section{Simulating many-body cell dynamics}\label{VM:ManyBodyCells}

Over the last two decades, scientific literature has been blooming with various means of simulating epithelial cell colonies. 
Each of these simulations can be separated by their respective efficiency (expressed in terms of consumed computational resources), the amount of cells/the size of tissues that can be simulated, the time scale of the simulated dynamics and the coarse grained level of precision. 
Choosing the right algorithm for the simulation of epithelial cells and tissues is a compromise between each of these key elements. 
Irrespective of the method, each algorithm includes part, or all, of the following features: 
short-range membrane-mediated attraction between cells, soft-core repulsion between cells, cell proliferation, cell death, cell motility, fluctuations, etc.
We will first give a non-exhaustive overview of commonly used modeling approaches for tissues at a mesoscopic level, giving a rough idea of the coarse-graining decisions made for every one of them.
Then we will dive into greater detail on how to implement a relaxation procedure according to the Vertex Model, refreshing aspects of the theoretical groundwork, describing required data structures and simulation steps and pointing out details of the simulation that can present pitfalls to a first-time implementation of the model.

\subsection{Modelling tissues at various levels of granularity}

When it comes to the representation of cells and tissues, there have been various different approaches trying to capture dynamics at different levels of detail. 
Of course there is the molecular level, where the objects of interest are proteins, their internal dynamics and their interactions with other parts of the system.
This level of detail can be tackled with the already established methods of MD simulations and other atomistic approaches.

At the other end of the scale, there are the macroscopic models, where a tissue is described via its material properties like the Young's modulus and its viscosity.
The latter can be exploited because large colonies of cells behave like viscous fluids, opening the the door to the shift to a more classical approach, where known techniques from material sciences and fluid dynamics can be applied to abstract away the processes relevant on a atomistic nano-scale. 

In this chapter, we will be focusing on the description of tissues on the mesoscale, where the overall scale of the system is still comparable to that of an individual cell but the dynamics do concern more than just individual cells.
In \cref{fig:VM:model_overview}, we provide a visual overview for a certain set of cell models on this scale located between the sub-cellular and super-cellular scale in terms of their measures for capturing the dynamics of the individual actors of their systems.

\subsubsection{Cellular Potts models}

On the sub-cellular end of the scale, there are the \emph{Cellular Potts} models, which build on the techniques employed in Ising-Models to describe cell dynamics on a discrete grid. 
A grid-cell can be assigned to a cell and a cell is represented by the entirety of cells assigned to it. 
This approach allows for the investigation of how cells change their shape depending on parameters of interfacial cell-cell interaction in competition with internal cell processes depending.
Further degrees of detail concerning cell polarity as well as interaction with the medium can be modeled using an appropriate Hamilton operator with the full time-evolution dynamics usually being simulated employing well-known Metropolitan-algorithm techniques. 

\subsubsection{Phase Field models}

Contrary to the grid-based approach of Cellular Potts models, the \emph{Phase Field} models use a description of individual cells via a phase field $\phi_i$, that takes on a value of $1$ inside and $0$ outside of cell. 
Between these two domains is the gradual transition corresponding to the interface or membrane of each individual cell. 
Combining the different phase-field functions for all cells in the system, one can again combine the overall interaction and internal energies into a hamiltonian and simulate the dynamics of the combined tissue via classical numerical techniques. 
This approach introduces more degrees of freedom to each individual cell than the cellular Potts model with the Phase fields potentially overlapping and the grid used for the numerical simulations being more fine-grained than that employed in the Cellular Potts model, but effectively both of these approaches work on a similar level of simulation.

\subsubsection{Active Network/Vertex/Voronoi models}

This class of models lies in the middle of our scale of cell detail with increasingly many simplifications and abstractions being introduced. 
In these \emph{Active Network} models, one tries to describe cell dynamics in terms of Network-interactions either by employing a simplified representation of a cell as a polygon and using the vertices of the polygons as the degrees of freedom (\emph{Vertex} Models) or by representing the cell by one or two points from which the cell shape and tissue topology can then be reconstructed (\emph{Voronoi} Models).
The interactions of cells are then described in terms of these degrees of freedom, with certain shape aspects still being retained, while the degrees of freedom per cell are reduced drastically compared to the prior two models. 

\subsubsection{Particle models}

The trend towards a higher level of abstraction continues with the \emph{Particle} models, where cells are represented by point particles with different kinds of interaction (i.e. adhesion forces at short distances and repulsion in the medium distance regime). 
This approach is known from material science of granular matter and allows for the simulation of way larger systems in terms of size with respect to the number of cells being simulated as we need to account for fewer degrees of freedom per cell. 
This class of models is the first where we see a loss of information concerning cell shape.
Not only have we simplified the shapes that we allow for cells to take (as in the Active Network class of models) but instead we have completely abstracted that aspect of the cells away, with only general pairwise interaction forces still being considered to describe cell-cell interactions.

\subsubsection{Continuum models}

As the ultimate consequence of the trend towards abstracting away degrees of freedom of each cell, one eventually arrives at \emph{Continuum} models, where the individual cell is no longer modeled directly.
It is replaced by a density distribution describing the distribution of cells in the system, which can then be used to calculate total system energies, which allow for predictions of overall colony/tissue evolution.
The description of whole cells via a density distribution is similar to the approach in the Phase Field models, where the individual parts of a cell were abstracted away into a density distribution/phase field describing the positioning of the cell in the system. 
The Continuum models then reach the next level of abstraction with the cells themselves being the parts no longer being considered individually but as an effective agglomerate whose dynamics are to be investigated.


\subsection{Bridging scales with adaptive numerical simulations}

Summing up the different approaches to tissue simulation at a mesoscopic level, one faces a trade-off forcing us to choose between the degrees of freedom for each cell or coarse-grained descriptions, the complexity and level of detail of or abstract effective interactions all while considering the need for computational resources to simulate the desired dynamics. 
This makes it hard to relate low-level cell properties to the features of large scale tissues. 
Therefore, in this chapter we will focus on a tissue description in the middle of the mesoscopic scale which still retains some degrees of freedom in terms of shape/membrane information while abstracting away details of complex interfacial cell-cell interaction: The Vertex Model. 
This model allows us to bridge sub-cellular and super-cellular scales, from cell level dynamics to tissue morphogenesis and growth.


\section{Simulation of tissue relaxation with the Vertex Model (VM)}

\subsection{Model details}

    Just like many other models for describing multi-cell dynamics, the Vertex Model (VM) can in principle be used to describe expanding tissues.
    Instead, due to its more common application to structural optimization in tissues, we will focus on that approach here and lay out its application in the scenario of tissue relaxation processes.
    
\subsubsection{History of the Vertex Model}

    Compared to other models, the the Vertex Model (VM) is a rather young introduction to the field of Biophysics. 
    Starting in 2007, the VM in its most common form has been introduced as good model to describe the structural properties of a tissue caused by its membrane dynamics through apical junctions. \cite{farhadifar2007influence} 
    This is a departure from other models trying to model the cell-level tissue dynamics like the Dissipative Particle Dynamics (DPD) approach\cite{liu2015dissipative}, which mainly focuses on the idea of the tissue dynamics being governed by the cells' nuclei. 
    There, the membranes were simply an intermediate structure reconstructed from the nuclei positions to determine, which cells neighbor each other and its dynamics can be simplified to the representation via two points for each cell.
    In years following the introduction of the VM, extensive theoretical and in-silico work on the VM by the group of Lisa Manning has shown that the it can predict a certain type of rigidity phase transition in tissues in terms of only one effective parameter, the \emph{isoperimetric ratio} $I=P_0/\sqrt{A_0}$ defined via the cells' target area $A_0$ and target perimeter $P_0$. \cite{bi2015density}
    A plethora of different analyses and implementations has emerged since then, both for expanding tissues as well as for the optimization/relaxation of tissues with a fixed number of cells. \cite{alert2021living,alert2020physical}
    
    We will focus on a very basic description of an optimization routine for the basic VM in two dimensions, which can easily be extended to account for even more constraints/interactions and higher spatial dimensions or it could be adapted to allow for the simulation of the time-dependent growth of tissues.

\subsubsection{The physical foundation of the VM as an energy functional}

    In the Vertex Model, one considers a tissue to be a collection of cells. 
    Each cell is here represented by a shape with a well defined area $A$ and a perimeter $P$ with the latter being a model description for the cell membrane, confining the cytoplasm spread across the cell area within.
    One does not usually keep track of a representation of the cell nucleus within this shape, especially not in the most simple form of the model. 
    Depending on how complex the description of each cell is supposed to be, one has many options to describe its shape, but the simplest one is certainly the representation via a polygon -- an oriented collection of points or vertices with edges in between defining the position of the cell's membrane.
    Then, from the position of the vertices, both the perimeter and the area of the cell can be calculated.
    
    Based on the area $A$ and the perimeter $P$ of a cell, the VM assigns a certain energy to its configuration that is intended to describe a set of counteracting processes favoring either an extended or reduced length of the cell membrane.
    This energy functional can -- in a simplified form, neglecting a constant coefficient -- be written as:
    \begin{align}
        E_{cell}(A,P) = (A-A_0)^2+ \frac{(P-P_0)^2}{r}. \hspace{30pt} \text{(see chapter on VM for detailed explanation)} \label{eq:VM:main_energy_functional}
    \end{align}
    In this, $A_0$ describes a certain target size the cell is trying to attain and $P_0$ is an effective target perimeter arising from counteracting forces of cell-cell-adhesion and contractile forces within the cell membrane of each individual cell. 
    Additionally, $r$ is a constant controlling the relative magnitude of the energies associated with cell size and cell perimeter.
    
    The VM does not need to reconstruct cell shapes and neighbor relations from simplified representations but instead it keeps track of the tissue topology via shared vertices between neighboring cells.
    Overall, each vertex in the simplest model of a confluent tissue (i.e. fully covering the surface) has three adjacent cells and is affected by their combined dynamics.
    Let us by $\vec{r}_i$ denote the position of vertex $i$ and by $\nabla_i$ the spatial gradient according to that position. 
    The detailed forces $F_{i,k}$ acting on a vertex $i$ within a cell $k$ can easily be calculated from \cref{eq:VM:main_energy_functional} by calculating the energy-gradient of that cell:
    \begin{align}
        \vec{F}_{i,k} = \nabla_i E_{cell}(A_k,P_k) = 2(A_k-A_0)\nabla_i A_k+ \frac{2(P_k-P_0)\nabla_i P_k}{r}. \label{eq:VM:energy_forces}
    \end{align}
    As both the area $A$ and the perimeter $P$ can be expressed analytically in terms of the positions of the vertices in the cell $k$, we are able to derive a precise analytic formula to link the positions $\vec{r}_i$ of vertices to the forces experienced by each vertex (see ex.~\ref{ex:VM:analytics_energy_functional}).
    The total force acting on a vertex is then obtained by summing up the contribution from all its neighboring cells.
    
    Once the energy gradients (or forces) at each vertex are known, we can then proceed to try and find local and global minima in the energy landscape of a tissue via gradient descent.
    \begin{example}

\label{ex:VM:analytics_energy_functional} 
{\bf Analytics of the energy functional.}
We want to develop the analytical expression for the gradient of the area $\nabla_i A$ and the gradient of the perimeter $\nabla_i P$ of a cell required for the analytical expression of \cref{eq:VM:energy_forces}.

Let $\vec{r}_i$, $\vec{r}_{i+1}$ and $\vec{r}_{i+2}$ be the positions of the vertices of a triangle.
Then the oriented area of the triangle can be calculated via the two-dimensional cross-product: $\vec{a}\times \vec{b}$
\begin{equation}
    \vec{a}\times \vec{b} = 
    \begin{bmatrix}
           a_1 \\
           a_2 
         \end{bmatrix} \times
          \begin{bmatrix}
           b_1 \\
           b_2 
         \end{bmatrix} =  a_1 \cdot b_2 - a_2\cdot b_1
\end{equation}
such that except for maybe the sign $A = \left(\vec{r}_{i+1}-\vec{r}_{i}\right)\times \left(\vec{r}_{i+2}-\vec{r}_{i}\right)$. Here, the sign of the area depends on the orientation of the three vertices in (counter-)clockwise orientation.

\textbf{ a.} Use the oriented formula for the area above to find an expression for the full area $A$ up to a correction of the sign of a polygon with vertices at $\vec{r}_1$, $\vec{r}_{2}$, $\ldots$ $\vec{r}_{n}$. \\
\textit{ [Hint: Starting from one point, cut the polygon in triangles and apply the formula for the triangle.]}

\textbf{ b.} Use the formula for the area of a full polygon to calculate $\nabla_i A$.\\
\textit{ [Hint: Consider how best to slice the polygon into triangles such that only very few triangles involve the point $\vec{r}_i$.]}

The length of an edge $l_{i,i+1}$ can be calculated via $l_{i,i+1} = \left| \vec{r}_{i+1}-\vec{r}_{i} \right| = \sqrt{(x_{i+1}-x_i)^2 + (y_{i+1}-x_{i+1})^2}$.

\textbf{ c.} Use the above formula to derive a sum expression for the full polygon perimeter $P$.

\textbf{ d.} Use the above formula to calculate the gradients $\nabla_i l_{i,i+1}$ and $\nabla_{i+1} l_{i,i+1}$.

\textbf{ e.} Derive the formula for $\nabla_i P$.

\end{example}
    
\subsubsection{Tissue boundary conditions and system size}
    
    To fully describe a tissue, one needs to consider how to deal with the boundary conditions at the tissue edge. 
    In a growing tissue simulation, one would opt for a finite size tissue growing on an infinite plain with a optionally a special treatment being applied to the cells being detected to be at the edge of an expanding tissue. 
    For the VM in its relaxation implementation, we will instead consider a \emph{confluent} or \emph{covering} tissue, i.e. we define a box with periodic boundary conditions (PBC) such that cells cover the entire box and periodic copies of the box tile the plain. 
    Hence the "outer cells" to the left and to the top of the box will share edges with the outer right and lower bottom ones respectively. 
    
    In this PBC scenario, it is crucial to have a reasonable minimum number of cells in the system to avoid the observation of finite size effects due to cells interacting with their own copies. 
    Such effects have been observed in various other applications like MD particle dynamics and can either be corrected for analytically or by scaling the system up. \cite{yeh2004system} 
    Here we face two competing effects: On the one hand, we want the system to be reasonably large to avoid finite size effects, on the other hand, the nature of the equilibration protocol will lead to quickly increasing relaxation simulation times with increasing numbers of cells. 
    Depending on how many equilibrations need to be run up until convergence, one may either decide to prioritize one or the other. 
    Generally we often limit ourselves to tissues of sizes below $100$ cells, as other groups have been forced to do, too.\cite{bi2015density}

\subsection{Implementing the VM relaxation procedure}
    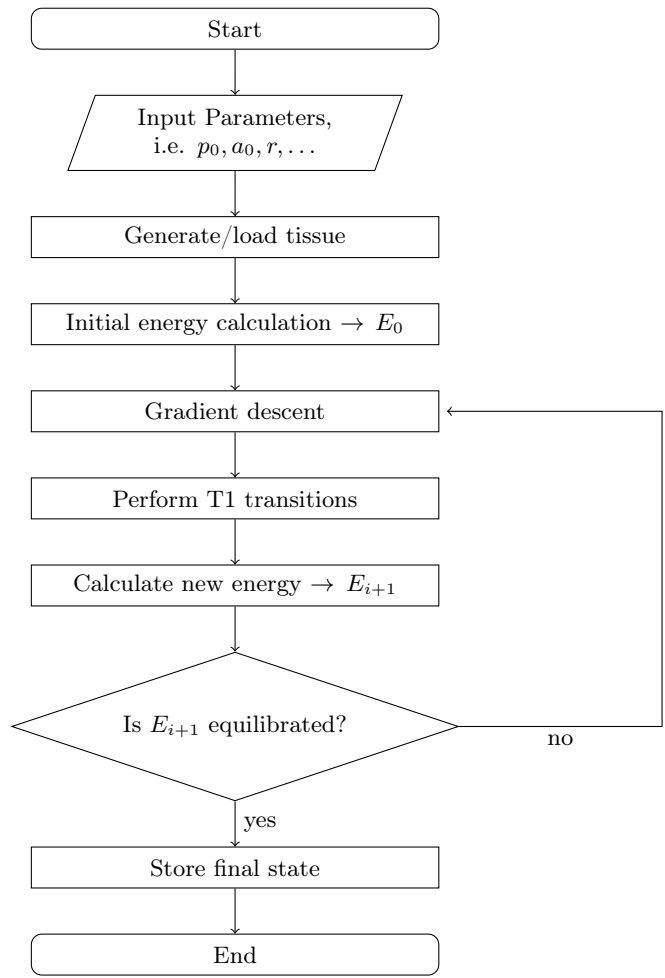
\begin{figure}[htp]
        \centering
        \begin{tikzpicture}[%
        start chain=going below,        
        node distance=6mm and 60mm,    
        every join/.style={norm},       
        ]
        \tikzset{
          base/.style={draw, on chain, on grid, align=center, minimum height=4ex},
          process/.style={base, rectangle, text width=16em},
          io/.style={base, trapezium, trapezium left angle=70, trapezium right angle=110, minimum width=3cm, minimum height=1cm, text centered, text width=3cm, draw=black},
          decision/.style={base, diamond, aspect=3, text width=13em},
          startstop/.style={process, rounded corners},
          coord/.style={coordinate, on chain, on grid, node distance=6mm and 25mm},
          nmark/.style={draw, cyan, circle, font={\sffamily\bfseries}},
          norm/.style={->, draw, lcnorm},
          free/.style={->, draw, lcfree},
          cong/.style={->, draw, lccong},
          it/.style={font={\small\itshape}}
        }
        \node (start) [startstop] {Start};
        \node (options) [io,join] {Input Parameters, i.e. $p_0, a_0, r,\ldots$};
        \node (generation) [process, join] {Generate/load tissue};
        \node (setup) [process, join] {Initial energy calculation $\to E_0$};
        \node (equilibration) [process, join] {Gradient descent};
        \node (T1_transition) [process, join] {Perform T1 transitions};
        \node (E_recalc) [process, join] {Calculate new energy $\to E_{i+1}$};
        \node (main_sim_switch) [decision, join] {Is $E_{i+1}$  equilibrated?};
        \node (finalization) [process] {Store final state};
        \node (end) [startstop, join] {End};
        \draw [->, shorten >=1mm]
                (main_sim_switch.east)  --node[anchor=north] {no} ++(27mm,0) 
                |-  (equilibration);
        \draw [->] (main_sim_switch) -- node[anchor=west] {yes} (finalization);
        \end{tikzpicture}
        \caption{\textbf{Process flow chart of the main simulation process of the Vertex Model.}
        It consists of initial setup steps to obtain a base tissue configuration, an equilibration loop being run until convergence is reached and a final step to store the resulting tissue configuration.}
        \label{fig:VM:main_flowchart}
    \end{figure}
    
    \subsubsection{Data structures for simulation}
    For the VM simulation, we require a set of data structures to represent the tissue and its cells as well as boundary conditions.
    
    We will call a data structure with two floating point coordinates a vector. It represents the two-dimensional position of a point in the plain.
    
    A cell is then described as a list of vectors representing the vertices of its polygonal shape. 
    Additionally, an integer index is associated with each of the vertices.
    This index is used to identify which vertices of different cells are meant to represent the same point in space.
    As each vertex is shared between three cells, the same index should occur exactly three times across the whole tissue. 
    
    A tissue is then represented by a list of cells as detailed before. 
    
    \subsubsection{Outline of the relaxation procedure}
    
    The overall VM simulation can be abstracted to a small set of main steps as presented in \cref{fig:VM:main_flowchart}.
    We want to outline the overall procedure roughly to give an idea of the relation between the steps as well as their function before going into more detail on their individual implementations.

    First, the optimization/relaxation procedure requires a certain set of input parameters for the process -- most notably the model parameters $A_0$, $P_0$ and $r$ -- as well as an initial tissue state to work on.
    For this initial state, it will either load a tissue configuration or generate it based on some settings, e.g. the number of cells to be considered and their average area.
    Then some initial setup computations are carried out including the calculation of the initial total energy $E_0$ of the tissue to ensure that a state change only ever reduces total energy with each optimization step.

    Then the actual optimization/relaxation starts.
    The optimization is run in multiple iterations with the $i$'th iteration starting with the energy $E_i$ and ending with the energy $E_{i+1}$ until we consider the total energy to have converged to a minimum.
    Each iteration involves only two major modification steps required to describe the tissue dynamics being considered.
    
    First, we have the energy minimization using gradient descent in the energy landscape described by the typical per-cell energy functional detailed in \cref{eq:VM:main_energy_functional} for a fixed tissue topology -- i.e. fixing which cell has shared edges with which other cell.
    In this, the individual vertices are moved around to reduce the total energy of the tissue without changing the topology, i.e. the number of vertices per cell or which vertices are connected by edges.
    This step can also be referred to as the equilibration step, driving the system to the closest (local) energy minimum and thus a stable equilibrium given the fixed topology.
    
    Then there is a second step in which the topology of the tissue may be updated in an operation referred to as a \emph{T1 transition}. 
    A T1 transition is the process of an initial shared edge between two neighboring cells being replaced by a shared edge between the two common neighbors of those cells.
    This allows for cells to lose or gain neighbors, which is the only way in which the tissue can change its structure.
    In the process of overall energy optimization, this change in topology can be quite a significant contribution on top of the equilibration step.
    Again, as in the equilibration step, a T1 transition will only be performed in case the total tissue energy after the transition is reduced relative to the energy at its start.
    
    After the performance of equilibration and T1 steps, the resulting energy $E_{i+1}$ of the tissue is calculated and compared to the previous energy $E_{i}$ at the beginning of the iteration. 
    If the change $\Delta E = |E_{i+1}-E_{i}|$ is above either an absolute or relative threshold, then the optimization step is repeated, starting from the equilibration step.
    Otherwise, the relaxation is considered to have converged and the final optimized state is stored. 
    
    \subsection{The individual steps of the VM relaxation procedure} 
    
    \subsubsection{Generation/Loading of initial state}\label{sec:VM:gen_loading_tissue}
    
    In order to relax a tissue, we need an initial guess for a state from which the relaxation can start. 
    
    \textbf{How to store a tissue configuration:} On the one hand, the state can be loaded from file, if it has been generated previously. 
    If you have no prior experience with reading and writing binary files, we recommend a textual representation of the tissue's state. 
    The tissue can for example be represented as follows:
    \begin{itemize}
        \item First, the PBC information, usually amounting to the vectors of the bottom left and the upper right corner of the bounding box if it is a rectangle.
        \item Then the actual tissue information is provided:
        \begin{itemize}
            \item The integer number $n_v$ of distinct vertices in the tissue, followed by $n_v$ vertices represented by two coordinates each.
            \item The number of cells $n_c$ of the tissue followed by $n_c$ representations of a single cell:
            \begin{itemize}
                \item First the number $m_v$ of vertices the cell's polygon shape has followed by $m_v$ entries for the vertices.
                \item Each of the vertices is represented by two integer indices $i_{pos}$ and $i_{pbc}$. 
                    The index $0\leq i_{pos} < n_v$ is the index of the vector in the initial list of $n_v$ vectors representing the position of the vertex.
                    The index $0\leq i_{pbc}$ on the other hand identifies the vertex in the topology. If two vertices have the same $i_{pbc}$ then they represent different copies of the same vertex in its $3$ adjacent cells.
            \end{itemize}
        \end{itemize}
    \end{itemize}
    All programming languages provide the tools to read such a textual format and due to its human-readable nature, it is also way easier to check for errors.
    Here, we have gone for a split between the positional information (the initial list of vector coordinates) and the description of the tissue topology (the list of cells and their vertices).
    As an alternative, one may merge the two sets of data and instead of storing $i_{pos}$ for a vertex in the cell directly store the vertex coordinates.
    One would still need to retain $i_{pbc}$, due to shared nature of vertices in a confluent tissue with periodic boundary conditions (PBC).

    \begin{figure}[htp]
        \centering
        \includegraphics[width=.9\textwidth]{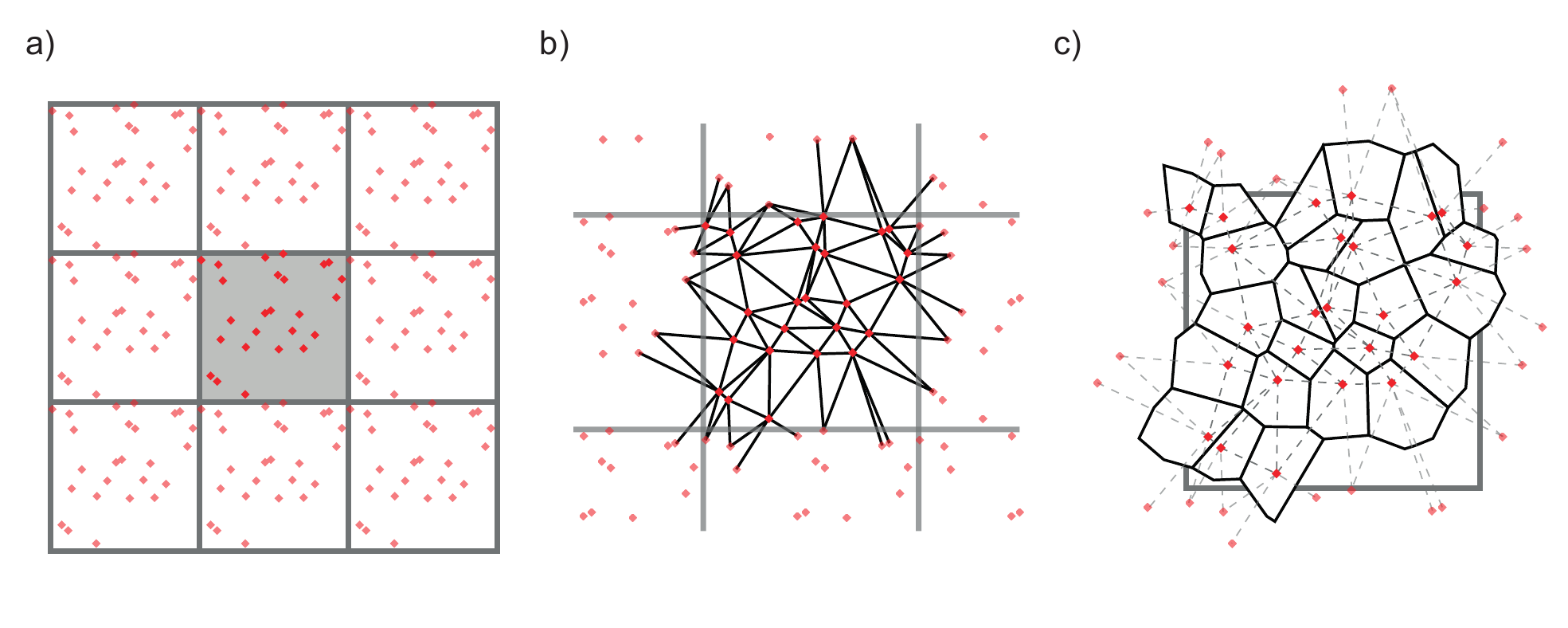}
        \caption{
        \textbf{Scheme of tissue generation.} 
        \textbf{a} Generation of randomly distributed points within the main PBC cell (grey) and their copies in the $8$ immediately neighboring PBC cells. At the end of generation, one polygon/cell will be associated with each point. 
        \textbf{b} Triangulation of the the set of points using the Delaunay-algorithm \cite{delaunay1934sphere}. An edge is drawn between two points if the respectively generated cells will touch in the end.
        \textbf{c} Construction of the Voronoi cells\cite{voronoi1908nouvelles} from the intermediate Delaunay-triangulation obtained in step \textbf{b}. The Voronoi cells constructed for the initially generated points are used as an initial set of cells for the relaxation process.
	}
        \label{fig:VM:pbc_generation}
    \end{figure}

    \textbf{How to generate a tissue:} If one does not have a previously generated tissue configuration, an initial state needs to be generated from scratch.
    The process for doing so in our simulator is visualized in \cref{fig:VM:pbc_generation}.
    
    First, we use a random number generator to obtain the coordinates of random points within the PBC box.
    For each of these points, we want to generate a corresponding cell in our tissue. 
    To account for the PBC, we first copy the generated points to its first $8$ neighboring positions (see \cref{fig:VM:pbc_generation}a) so that we can account for all neighbor relations between cells in our system. 
    
    Then we construct the Delaunay triangulation \footnote{The Delaunay triangulation\cite{delaunay1934sphere} is a complex algorithm generating a set of edges between points fulfilling certain conditions concerning the position of points being outside of the circumcircles of the constructed triangles as well as conditions on the angles between edges. It is mainly used as an intermediate step for the generation of the cell shapes obtained by constructing the Voronoi cells from the triangulation. An edge in the triangulation indicates that two cells will be neighbors at the end of the generation process. See also \url{https://en.wikipedia.org/wiki/Delaunay_triangulation}} of all the points in these $9$ PBC cells to identify, which cells will eventually be sharing a common edge (see \cref{fig:VM:pbc_generation}b).
    
    Finally, from the intermediate Delaunay triangulation, we construct the actual shape of the cells by constructing the Voronoi cells\footnote{A Voronoi cell\cite{voronoi1908nouvelles} of a point is the set of points in the plane closer to this point than to any other point. It can be constructed using the perpendicular bisectors of the edges of the Delaunay triangulation. See also \url{https://en.wikipedia.org/wiki/Voronoi_diagram}} of each point and eventually only retaining the Voronoi cells of the points within the main PBC cell (see \cref{fig:VM:pbc_generation}c). 
    This leaves us with a collection of cells that we can use as a basis for our relaxation procedure. 
    
    The way the initial points are distributed within the PBC box can lead to different initial configurations.
    Still, assuming that the relaxation procedure is efficient and stable enough, the VM optimization procedure should still eventually lead to statistically comparable results no matter how the cells were seeded initially.
    The transient evolution during the optimization will probably be different, though.
    
    As per usual, there are other options for the generation of initial configurations depending on which properties you are trying to investigate. 
    Tiling the PBC with a repeating pattern of cells with identical shapes is an option, but the nature of the VM makes it very unlikely for the degree of symmetry of the system to drop during tissue relaxation. The statistical results when starting from such a configuration may thus not be generalizable.
    We therefore recommend starting from a more random initial condition. 
    
    \subsubsection{Energy calculation}
    
    The energy of the system is calculated by looping over all cells, calculating their perimeter and their cell area and -- based on that -- the cell energy via \cref{eq:VM:main_energy_functional}.
    By adding up those energies for all cells, we obtain the full-tissue energy.
     
    Be aware that you may need to account for the periodic boundary conditions to obtain absolute positions for vertices when calculating $P$ and $A$ for a cell depending on how you decide to deal with the PBC.
    Accidentally using a periodic copy of a vertex for the calculation can lead to very large perimeters and negative or very large areas. 
    Pay attention to the individual cells' values during the calculation to identify such issues of periodicity. 
    A useful tool here is finding the closest periodic copy of a vector given the PBC of the box, which is a known problem in vector algebra and will not be detailed here.
    
    \begin{figure}[htp]
        \centering
        \includegraphics[width=.9\textwidth]{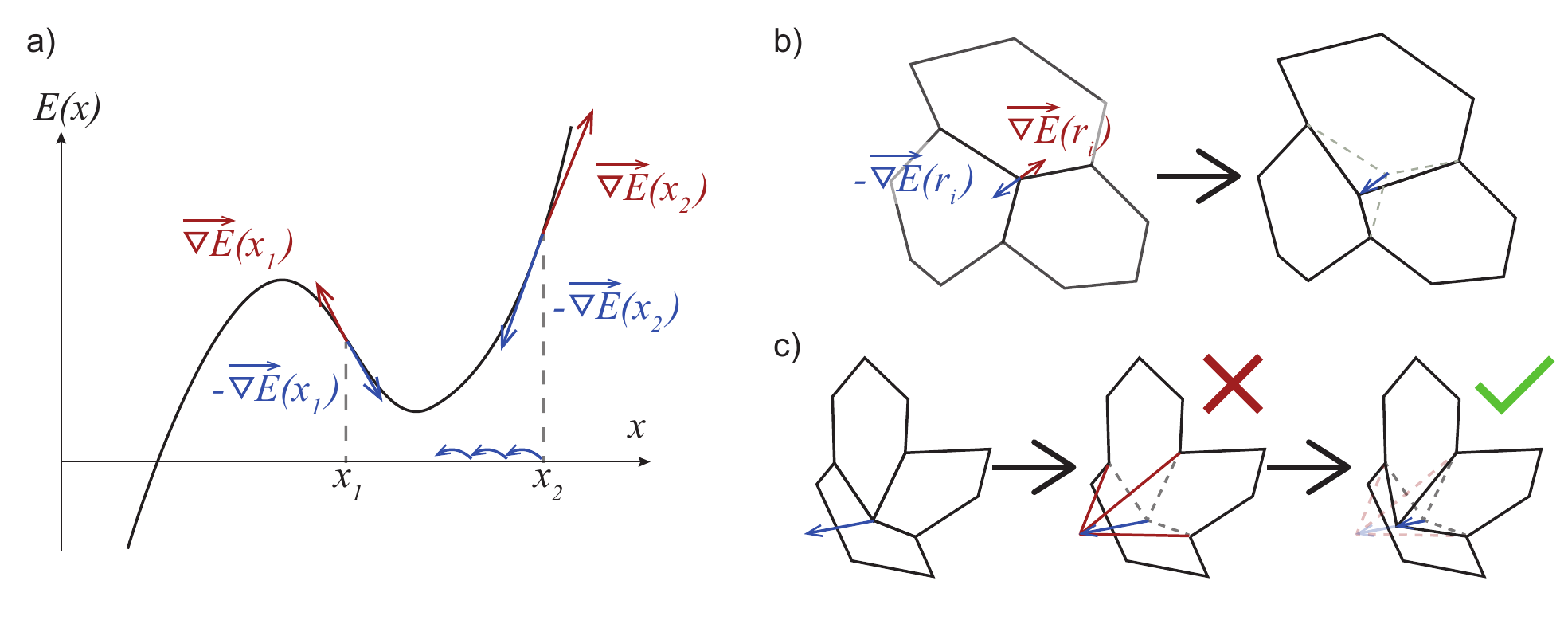}
        \caption{\textbf{Gradient descent in tissues.}
        \textbf{a} General visualization of the concept of the gradient descent. Starting at either position $x_1$ or $x_2$, the gradient $\vec{\nabla}E$ (red) will always point towards the highest increase of $E$. One can therefore follow the direction of $-\vec{\nabla}E$ (blue) towards a (local) minimum. This may need to be repeated several times to actually reach the desired minimum.
        \textbf{b} Process of gradient descent for a vertex in the tissue. After calculating $-\vec{\nabla}E$ for a vertex from the contribution of all three neighboring cells, the vertex is moved along $-\vec{\nabla}E$, which changes the length of cell edges or rather the length of cell membranes $P$ as well as cell areas $A$. 
        \textbf{c} Possible issues with geometric constraints. When following the negative gradient $-\vec{\nabla}E$ for a vertex, it may happen that the vertex crosses another edge of one of its three adjacent cells, thus creating intersecting pairs edges amounting to intersecting cell membranes. This is an undesired effect and the gradient descent step needs to be repeated with a shorter step length to try and avoid this from happening.}
        \label{fig:VM:gradient_descent}
    \end{figure}
    \subsubsection{Equilibration/Gradient descent step}
    
    In the equilibration step, the tissue is supposed to be optimized with respect to the energy functional provided in \cref{eq:VM:main_energy_functional}.
    For this purpose, we can sum up the forces acting on the individual vertices resulting from the adjacent cells' energy functionals as derived in \cref{eq:VM:energy_forces}.
    This assigns the energy gradient $\nabla_i E$ to each vertex $i$ in the tissue, pointing towards the direction of steepest increase of energy (see \cref{fig:VM:gradient_descent}a).
    We can then follow the opposite direction of this gradient for individual vertices to converge towards the point of lowest energy (see \cref{fig:VM:gradient_descent}b). 
    As we do not strive for the simulation of an accurate time evolution but instead want to explore the configuration space for ideal tissue states, we can either perform gradient descent steps on each vertex individually or on all vertices in parallel. 
    Also we have the option to repeatedly perform the gradient descent steps until the energy cannot be optimized any further or just once. 
    If the step is repeated until convergence, we have reached a local minimum in tissue energy and a change in topology would be required to converge the tissue any further. 
    
    While performing the gradient descent, it is vital to account for constraints imposed on the cells, especially at high values of $P_0$.
    In scenarios with long target perimeters for a given target area, the energy of the tissue when making the full step might actually increase or cells may be driven towards non-convex shapes or even towards the edges of their polygonal shapes intersecting (see \cref{fig:VM:gradient_descent}c). 
    While it may not be an issue to have a cell shift towards a non-convex shape, intersecting edges -- amounting to intersecting cell membranes -- do not represent a (physically) reasonable configuration of the tissue.
    Hence such aspects need to be checked whenever a vertex has been moved and if the move has forced the tissue into an unreasonable state or lead to an increase in energy, the step needs to be reverted and a smaller step size should be attempted (see \cref{fig:VM:gradient_descent}c). 
    It can happen, that no stride length along the negative gradient allows for the constraints to be kept, this can for example be tested by repeatedly halving the step length and retrying again and again. 
    If several reductions in step length do still not allow for a decrease in energy while adhering to constraints, the step should be considered as failed and either another vertex should be attempted or the next operation should be executed instead.
    
    Instead of immediately moving a vertex along the entire length of its gradient vector, we often use a relative initial step length for our simulations and also recommend this to our readers.
    I.e. once the gradient $\nabla_i E$ has been calculated, try a first step of length $\alpha \cdot \nabla_i E$ with for example $\alpha = 10^{-1}$ or $\alpha = 10^{-2}$, depending on how fine-grained you want your relaxation to be.
 
    \begin{figure}[htp]
        \centering
        \includegraphics[width=.9\textwidth]{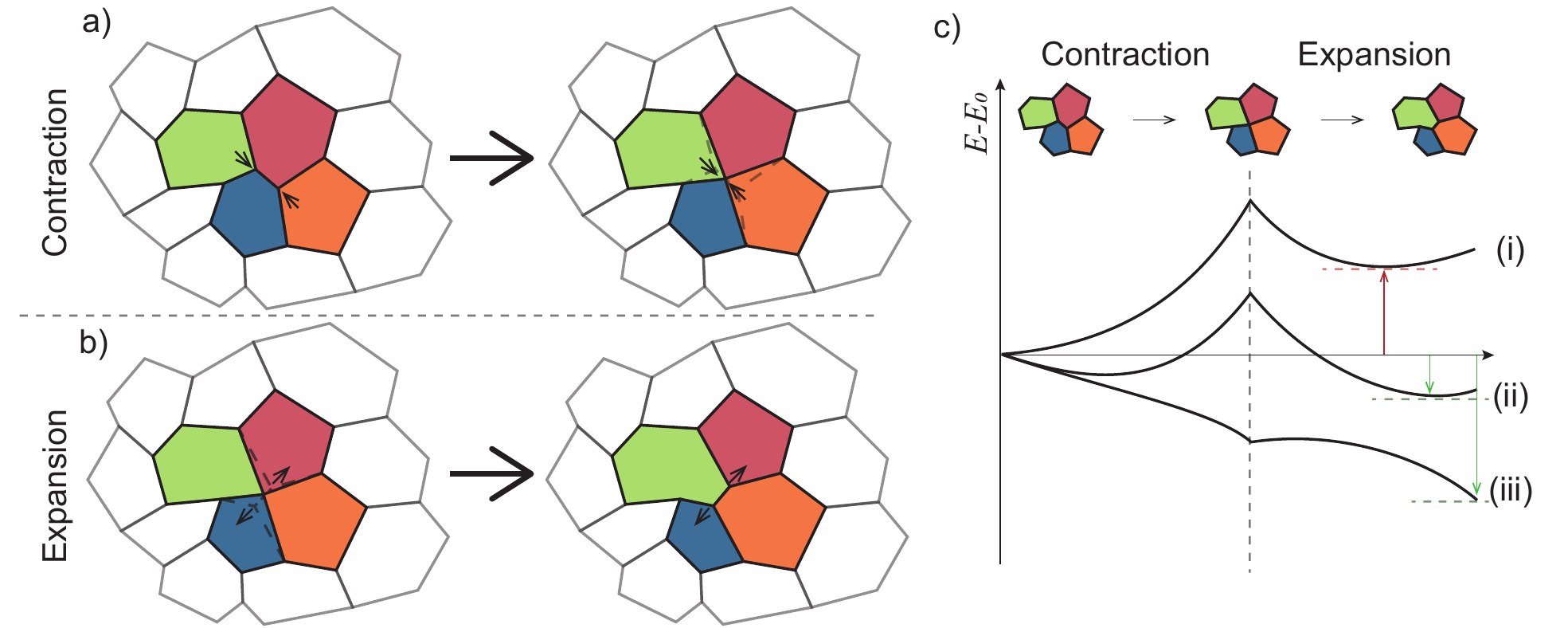}
        \caption{\textbf{Tissue topology changes via T1 transitions.}
        \textbf{a} Visualization of the \emph{contraction} step of the T1 transition, where an edge between two cells is collapsed into a point.
        \textbf{b} Visualization of the \emph{expansion} step of the T1 transition following the contraction in step \textbf{a}. A new edge is generated perpendicular to the original edge that has been collapsed and expanded. It is now a shared edge between the two common neighbors of the two cells whose initially shared edge has been collapsed. 
        \textbf{c} Examples of qualitative energy profiles observed during T1 transitions. Where \textbf{(i)} and \textbf{(ii)} do have an energy barrier towards the collapsed intermediate state, \textbf{(iii)} shows a decreasing energy profile towards the collapsed state. The latter is possible, if the collapsed state is actually energetically favorable but due to vertices only ever having three neighbors, the edge cannot fully collapse by gradient descent alone. The decision to accept or reject a T1 transition is then made based on the minimum energy observed during the expansion step. Here, \textbf{(i)} has a minimum expansion energy above the initial energy and is therefore rejected, whereas \textbf{(ii)} and \textbf{(iii)} have an energy minimum during expansion that is below the initial energy. Hence the T1 transition is accepted and the expansion state with minimum energy is picked as the post-transition state.}
        \label{fig:VM:T1_transitions_scheme}
    \end{figure}
    
    \subsubsection{T1-transition step}
    
    The gradient descent step in which the topology of the tissue is always maintained is then followed by a step, where the topology is modified in a physically and biologically reasonable manner while maintaining confluence of all cells, i.e. without holes in the tissue forming. 
    In real tissues, neighboring cells can detach from each other, which needs to be represented in this step via so-called \emph{T1-transitions}.
    A T1-transition is the process of two neighboring cells losing their shared edge and a new edge forming between their two common neighbors instead (visual representation in \cref{fig:VM:T1_transitions_scheme}).
    
    We identify possible candidates for a T1 transition via the length of the shared edge between two cells, i.e. by checking whether the edge length is below a certain length threshold length $l_T$. 
    If the gradient descent in the energy functional has driven an edge towards collapse therefore resulting in a short length $l<l_T$, then it might be energetically beneficial to replace it with an edge between two other cells instead, while maintaining the constraint that each vertex should have exactly $3$ adjacent cells. 
    The two cells between which the new edge is formed are the two cells that do not contain the relevant edge but do contain the vertices delimiting the edge. 
    As each vertex is a part of exactly three cells, these two are uniquely identifiable.
    Once all possible candidate edges with length $l<l_T$ have been identified, we successively try for every one of them to perform a T1 transition and then check if it has reduced the total energy of the tissue. 
    
    Performing a T1 transition involves first actually collapsing the candidate edge to a point from its original state associated with the energy $E_{T1,0}$ as visualized in \cref{fig:VM:T1_transitions_scheme}a. 
    This can either be done instantaneously or with several intermediate steps. 
    In a real tissue, this is a continuous operation, thus being a gradual contraction of the edge with the surrounding tissue adapting to the change.
    It is therefore reasonable to perform the collapse in several intermediate steps with tissue relaxation/gradient descent steps being performed for each of the intermediate steps to mimic local tissue response to the topology change. 
    
    Once the edge has been collapsed into a vertex now neighboring four cells, a new edge is created perpendicular to the original edge and expanded up to a maximum length of $l_E$ as visualized in \cref{fig:VM:T1_transitions_scheme}b.
    This is again performed in a gradual manner with relaxation of the tissue performed on the intermediate steps of the expansion. 
    One vertex in the collapsed state having $4$ neighboring cells is fine because it is only an intermediate configuration and is immediately transformed back into a valid state as soon as the expansion of the new edge starts. 
    During the multiple intermediate steps of edge expansion, for each length of the expanded edge $0< l\leq l_E$ the resulting relaxed total tissue energy $E_{T1}(l)$ is calculated and retained. 
    
    Once the full expansion has been performed, an energy profile depending on the expansion length $l$ similar to that in \cref{fig:VM:T1_transitions_scheme}c should have been observed. (The set of energy profiles in \cref{fig:VM:T1_transitions_scheme}c is not exhaustive and just meant to qualitatively illustrate how energy can evolve during the transition process).
    If at no point during the expansion an energy $E_{T1}(l) < E_{T1,0}$ has been observed, then the tested transition is not favorable to overall tissue energy and it should be rejected while reverting to the original pre-T1-state with energy $E_{T1,0}$.
    If instead during expansion at length $l_m>0$ a minimal energy $E_{T1}(l_m)= \min_l E_{T1}(l)$ below the initial energy before the transition (i.e. $E_{T1}(l_m) < E_{T1,0}$) has been observed, then the topology and tissue state associated with this length $l_m$ should be chosen for any further operations. 
    
    This is repeated until all possible candidate edges have either been rejected or collapsed.

    \emph{Some notes on the implementations of T1 transitions:} 
    \begin{itemize}
        \item The data manipulations required for performing the T1 transitions include the calculation of the position of the center point of the edge to which the edge collapses, the subsequent calculation of the end points of the new edge from the direction of the original edge with the new edge being perpendicular as well as the removal and addition of vertices to the respective cells involved in the T1 transition. 
        \item During the relaxation of the tissue performed during the intermediate steps of contraction and expansion, the positions of the vertices delimiting the old and the new edges respectively should not be relaxed along with other vertex positions. 
        \item The number of intermediate steps as well as the number of relaxation steps within each intermediate step influence the outcome of the overall VM optimization process.
        Fewer T1 transitions are accepted if no relaxation is performed, as the relaxation reduces overall tissue energy and makes it more likely for a state during expansion to be below the initial energy $E_{T1,0}$. 
        \item When the T1 transition moves from contraction to expansion, the topology of the tissue only needs to be updated once and only the four involved cells need to be updated but special care needs to be taken that the right vertices are part of the right cell after the transition.
        The first check should be that the correct cells have lost one vertex and the other two have gained one and then a visual inspection of the resulting tissue should be performed during testing to make sure the vertices are at the right position.
        This is made easier by keeping the vertices within each cell consistently in a (counter-)clockwise orientation because then the two vertices of the new expanding edge can be obtained by rotation of the original edge (counter-)clockwise and rescaling. 
        Keeping the same orientation of vertex orientation in cells and the rotation of the original cell then helps with assigning the correct new point to the correct cells and only requires visual inspection once or twice to make sure it is done correctly.
    \end{itemize}

    \subsubsection{Check for Convergence}
    
    Convergence of a series of energies $E_i>0$ is usually identified via the combination of an absolute and a relative threshold of change. 
    Let $\Delta E_i = \left|E_{i+1}-E_i\right|$ denote the absolute change and $\varepsilon_i = \frac{\Delta E_i}{E_i}$ the relative change of energy from iteration $i$ to iteration $i+1$.
    We consider the energy to be converged and therefore the tissue to have relaxed once $0\leq \Delta E < E_T$ is below an absolute threshold $E_T$ or $\varepsilon < \varepsilon_T$ is below a relative threshold $\varepsilon_T$.
    
    The absolute threshold may at first glance seem sufficient to identify very minor changes in the occurring energy values and thus convergence of the process, but there are situations where this is not a good measure for how far the optimization has come. 
    Indeed, if the optimum configuration is at a very high energy scale, then the absolute steps $\Delta E_i$ between iterations may stay at a large scale for a very long time without significant progress in terms of changes to the tissue configuration.
    The relative threshold is therefore a good measure for whenever we do not know the absolute scale of energies and want to estimate the chance for significant overall progress through the relaxation routine. 
    
    If the relative change is below e.g. $\varepsilon_T=10^{-3}$, we can be sure that independent of the absolute scale of the energy the achieved change does not significantly affect its scale anymore. 
    Furthermore the effect of any further relaxation iterations is also expected to be small as big changes tend to happen early on and successive iterations generally exhibit diminishing returns with only occasional spikes in $\Delta E_i$.
    Hence we can consider the system to be converged.

    If the absolute scale of the optimum energies is known precisely, the absolute threshold $E_T$ alone can provide us with a good measure of convergence. 
    In general, one should always consider combining the two for the most reliable identification of convergence.
    
    On top of that, it is always a good idea to limit the maximum number of relaxation iterations being performed to a reasonable number when implementing a simulator oneself. 
    This can help if there is an issue with the relaxation that has not originally been considered but we do not want to lose the progress that we have already made, for example when a large system has been processed for quite a long time.

    \subsubsection{Storage of tissue state}
    
    For the storage of the final tissue state, we again recommend using a textual description as presented for the Generation/Loading step or -- for more advanced readers --  going for a binary (i.e. machine-readable) custom format containing the same information. 
    The latter has the advantage of being more storage-efficient but the downside of being non-human-readable so if there is an issue, it will be harder to debug. 
    
    For debugging purposes, you may also consider outputting a visual representation in the form of an SVG-file.
    SVG is a vector image format ideally suited to output polygonal shapes like the ones used here.
    Additionally, SVG is written in a human-readable text description of where to put vertices of polygonal shapes. 
    There are various tutorials for outputting SVG format figures,\footnote{See for example \url{https://developer.mozilla.org/en-US/docs/Web/SVG/Tutorial}} and we would recommend following any one of them as in our experience, the visual representation of the tissue state along the relaxation process has helped tremendously with debugging.

\subsection{Technical and analytical limitations}
    
    The implementation of the VM as explained before is quite restricted in the possible system sizes that it can accommodate. Especially the complex relaxation process of the T1 transitions has a heavy toll in terms of computations as not only the number of candidates for T1 transitions increases but also the number of required relaxation steps during each T1 transition. 
    Hence an increase in the number of cells will very quickly increase the computation time required to fully optimize the tissue. 
    This has been observed by other groups as well\cite{bi2015density}, which have therefore restricted their analysis to double-digit numbers of cells.
    Reasonable statistics of observables can then be obtained mainly by re-running the relaxation for different random initial configurations with the same relaxation parameters and analyzing the different configurations independently.
    
    Another less obvious issue of the described relaxation process of the VM is the constraint imposed on the average cell area.
    Due to fixed PBC causing the overall area of the confluent tissue to be constant and no changes in the number of cells via apoptosis or proliferation, the average cell area does not change during the simulation. 
    A relaxation operation of the PBC could help mitigate the issue, but with the described methods, the parameters should be chosen carefully.
    
    While introducing cell apoptosis and proliferation may seem like the obvious solution to the above problem, we would like to point out some considerations for these processes that make it hard to bring them into the system without experimental data to answer these questions:
    \begin{enumerate}
        \item How is cell division triggered? Is it a purely random process or is there a condition on cell parameters? If the decision to split a cell is made based on its perimeter or area, this will alter certain statistics of the tissue. 
        \item Is there a prelude to cell division? When cells in real tissues divide, they will grow to a larger size before actually forming a separating membrane. It may therefore be a good idea to account for this growth by altering cell parameters $A_0$ and $P_0$ once the decision to divide has been made. 
        \item In which direction is a cell split in half during proliferation? This is relevant due to the possible effect on average cell elongation and other statistics relating to cell orientation/polarization. As a rule of thumb it might be wise to split a cell in half along its main axis that can be determined via principal component analysis (PCA).
        \item How is cell apoptosis triggered? Is it a purely random process or is it coupled to energy and/or size considerations?
        \item How do we distribute the area of a dying cell among its neighbors? 
        We cannot just remove it from the tissue without losing confluence. 
        Instead we need to account for its former neighbors coming into contact with each other once the cell is removed. 
        As we generally want vertices with only 3 neighboring cells, we cannot just collapse the cell to a point and be done with it, but instead we need to divide up its space and introduce a new topology for its neighborhood.
        How does one determine, which of the former neighbors will be sharing a new edge once the dying cell has been removed?
    \end{enumerate}
    
\subsection{Possible tweaks and optimizations}

    In the implementation of the gradient descent operation both in its own step and as a substep of the T1 transition's relaxation, there is the option to go for a stochastic approach to optimize the convergence of the tissue.
    In our experience, the various constraints of the individual vertices makes it inefficient to perform the gradient descent on all vertices simultaneously, because any constrained vertex will make it hard to relax all other vertices. 
    Picking one individual vertex and then relaxing it within the energy landscape defined by the remaining ones is therefore preferable. 
    Furthermore, to avoid the effect of always performing the gradient descent in the same order on all vertices, the order of vertices within the gradient steps can be chosen randomly prevent cyclic updates from happening that could just amount to a shift of the overall tissue but not generate any real new tissue relaxation effect.
    This arbitrary choice of order of optimization is possible due to the nature of our simulation not trying to adhere to a real time evolution but instead performing an exploration of phase space. 
    
    Whenever there is an operation run in parallel across the entire tissue like the generation of gradients, the determination of T1 edge candidates and the likes, parallelization can be considered. 
    Yet, a main issue is the sequential nature of certain steps like the intermediate relaxation of T1 transitions, which limits the ability to parallelize T1 transitions occurring on multiple edges at the same time.
    This is a main reason for the constraint on the number of cells that can be relaxed in a reasonable time span. 
    

\section{Conclusion}

On the scale of the mesoscopic tissue description, we have opted to detail an implementation of the Vertex Model (VM), a description technique, where aspects of cell shape are retained but simplified to a polygonal shape.
This places it firmly between more complex descrptions of cell shape like the Cell Potts and the Phase Field approach and more abstract techniques like the Particle and Continuum models. 

We have laid out the parts of the VM theoretical foundations necessary for the implementation, going into detail on some analytical aspects.
Building on that, we have laid out the necessary data structures for simulating a tissue in the VM formalism required to keep track of tissue topoloy. 
The detailed guide for the implementation of the individual steps for the use of the VM for tissue relaxation includes tips regarding pitfalls of implementation as well as possible extensions and optimizations of the simulation that the limited space of this section does not allow to go into detail on. 
The depicted algorithm is able to reproduce jamming behavior and tissue stratification previously described for in-silico investigations by other groups\cite{bi2015density} and can easily be extended to account for more recent developments of the field like nematic orientation behavior.

The interested reader may want to build upon the described algorithm to set up the simulation of a growing non-confluent tissue with cell proliferation and intermediate structure optimization/relaxation steps.
We recommend this as an excellent challenge to reconsider the choices made for the implementation of the VM presented here, e.g. the assumption of a confluent tissue which reduces the amount of edge cases one needs to consider in the implementation of T1 transitions and the analytical form of the energy functional. 


\section{Exercises}

\begin{figure}
    \centering
    \includegraphics[width=.8\textwidth]{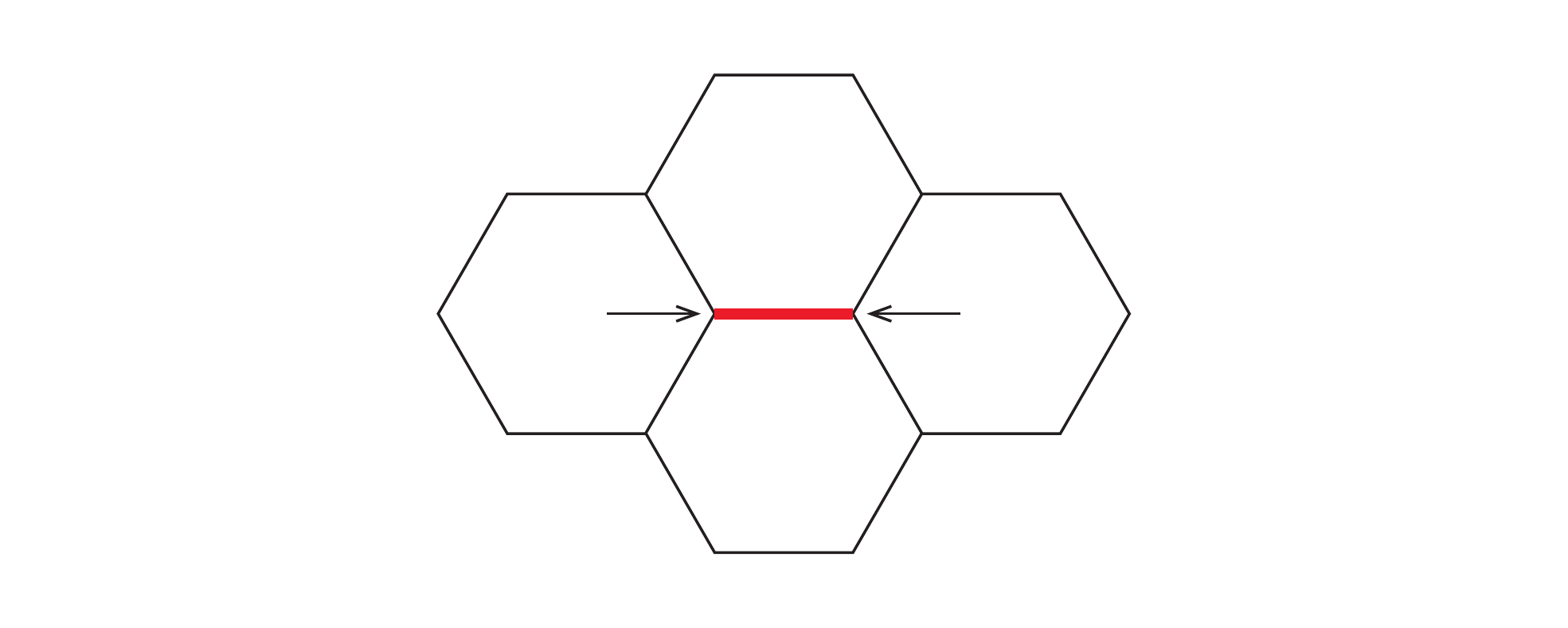}
    \caption{Simplified system for analyzing a T1 transition.}
    \label{fig:VM:toy_T1_system}
\end{figure}
\begin{problem}
\label{prob:VM:understanding_T1_transitions} 
{\bf Tissue topology and T1 transitions.}
The energy profile observed during a T1 transition as depicted in \cref{fig:VM:T1_transitions_scheme}c can be quite complex. 
It depends heavily on the choice of parameters $P_0$ and $A_0$ in \cref{eq:VM:main_energy_functional}, which influences both the optimum shape a cell can attain within the confluent tissue and how the energy functional during a transition evolves. 

\textbf{ a.} Convince yourself that, in a confluent tissue with periodic boundary conditions where all vertices have three adjacent cells, on average each cell must have $6$ vertices. \\
\textit{[Hint: Consider the angles at the vertices]}

\textbf{ b.} Let us now fix $A_0=1$ and only consider the influence of the choice of $P_0$. Find the optimum shape of a cell depending on $P_0$ by distributing the points used for generating the tissue in a grid pattern and then running the relaxation step until it converges. 
How does the choice of $P_0$ influence the shape? 
Does the shape or orientation of cells depend on how you chose the initial grid pattern? 
There should be a change in behavior of the cell shape depending on $P_0$. Where is that critical value of $P_0$?\\
\textit{[Hint: A circle is the shape with the shortest possible perimeter relative to the square root of its area]}

\textbf{ c.} Setup a simplified system of four regular hexagonal cells as depicted in \cref{fig:VM:toy_T1_system} and perform a T1 transition on its center edge. Perform this T1 transition with intermediate steps and keep track of the overall energy profile depending on $P_0$. 
How does the energy profile observed during the T1 transition change if you perform gradient descent/relaxation steps at each intermediate step of contraction and expansion?\\
\textit{[Note: You can also go for a system with periodic boundary conditions but then you will need to include more than these four cells. Also the results may depend on your chosen system size.]} \\

\end{problem}

\bibliography{MultiCellSimulation_chapter}

\end{document}